\documentclass[preprint2]{aastex}

\shorttitle{THE TILT OF THE FUNDAMENTAL PLANE}

\shortauthors{Trujillo, Burkert \& Bell}

\begin{document}

\title{THE TILT OF THE FUNDAMENTAL PLANE: THREE-QUARTERS  STRUCTURAL
NONHOMOLOGY, ONE-QUARTER STELLAR POPULATION}

\author{{I. Trujillo$^1$}, {A. Burkert$^{1,2}$} and Eric. F. {Bell}$^1$}
\affil{(1) Max--Planck--Institut f\"ur Astronomie, K\"onigstuhl 17, D--69117
Heidelberg, Germany\\
(2) Universit\"ats--Sternwarte M\"unchen, Wendelstein Observatory, Scheinerstr. 1, D-81679 Munich, Germany}
\email{trujillo@mpia-hd.mpg.de, burkert@usm.uni-muenchen.de, bell@mpia-hd.mpg.de}

\begin{abstract}

The variation of the mass-to-light ratios M/L of early type galaxies as
function of their luminosities L is investigated. It is shown that the tilt
$\beta=0.27$ (in the B--band) of the fundamental plane relation  $M/L\propto
L^{\beta}$ can be understood as a combination of two effects: about
one-quarter  (i.e. $\Delta \beta =0.07$) is a result of systematic variations
of the stellar population properties with increasing luminosity. The remaining
three-quarters (i.e. $\Delta \beta =0.2$) can be completely  attributed to
nonhomology effects that lead to a systematic change of the surface
brightness profiles with increasing luminosity. Consequently, the observed tilt
in the K--band ($\beta=0.17$) where stellar population effects are negligible,
is explained by nonhomology effects alone. After correcting for
nonhomology, the mean  value of the mass-to-light
ratio of elliptical galaxies  (M/L$_B$) is 7.1$\pm$2.8 (1$\sigma$ scatter).


\end{abstract}


\section{Introduction}

Traditionally, the global properties of elliptical galaxies have been
characterized by  three observables: the central velocity dispersion
$\sigma_0$, the effective or half-light radius $r_e$, and the mean surface
brightness within the effective radius $I_e$. It is generally assumed that
elliptical galaxies are relaxed particle systems. From a theoretical point of
view, the virial theorem then predicts a relationship between the gravitational
mass, the velocity dispersion and the radius. This can be written as a
relationship between $I_e$, $\sigma_0$ and $r_e$. These observables should then
be constrained to a two-dimensional plane in the above three-space. Elliptical
galaxies  are indeed found to lie along a plane; however, the observed plane
(popularly known as the fundamental plane [FP]; Djorgovski \& Davis 1987, Faber
et al. 1987, Dressler et al. 1987)  deviates from the prediction using a
constant $M/L$. This deviation is known as the ``tilt'' of the FP.  Since the
above prediction is based on the assumption that the mass--to--light ratio
and the global structure of the elliptical galaxies are not dependent on the
luminosity, the explanations for the tilt have been attributed to deviations
from one or both of these hypotheses.  

Following on that,  Faber et al. (1987) suggested that the  $M/L$ values
increase systematically with increasing luminosity. Using this interpretation,
the observed averaged value of the FP tilt over a large group of galaxy samples
in the B--band is written as $M/L\propto L^{\beta}$, with $\beta$=0.27$\pm$0.08
(Prugniel \& Simien 1996). Since brighter ellipticals are systematically
redder, at least part of the tilt must be understood as stellar population
variations along the different luminosities of elliptical galaxies. However,
the presence of a ``remaining'' tilt ($\beta$=0.17$\pm$0.01)  in the K--band 
FP (Pahre, Djorgovski \& de Carvalho 1998), where population variations are
weak, implies that this solution alone cannot explain completely the observed
slope. In fact, the variation in the properties of the  stellar population as
deduced from broad band colors and the Mg$_2$ index explain only one-third of
the slope, i.e., $\Delta \beta\sim0.10$ in the B-band (Tinsley 1978, Dressler
et al. 1987, Prugniel \& Simien 1996). We show this more explicitly in Figure \
1a, where the luminosity dependence of estimated stellar M/L ratios in $B$
(open circles) and $K$-bands (filled circles) is plotted for a sample of 911
morphologically selected early-type galaxies from the Sloan Digital Sky Survey
(SDSS) Early Data Release and the Two Mass All Sky  Surveys (2MASS) (Bell et
al.\ 2003). The dependence of stellar M/L ratios on galaxy luminosity is weak
in the $B$-band, in line with previous studies, and is negligible in the
$K$-band.  Taken together, the above is consistent with the conclusion that the
dominant part of the tilt of the FP ($\Delta \beta \sim 0.2$) is not a
population effect.

Changes in the mixture of dark-to-visible matter as function of luminosity
would  be a plausible explanation for the remaining tilt. However, 
observations (Romanowsky et al. 2003) demonstrate  that the amount of dark
matter within an effective radius is in general negligibly small. Furthermore,
systematic variations in kinematical structure (van der Marel 1991, Bender,
Burstein and Faber 1992, hereafter BBF)  and radial anisotropy (Ciotti, Lanzoni
\& Renzini 1996, Ciotti \& Lanzoni 1997)  cannot produce the observed tilt. The
role played by rotation is also small (Busarello et al. 1997).

In this Letter, we argue that the remaining tilt of the FP is due to a
systematic variation of the structural (and its associated dynamical)
nonhomology of the elliptical galaxies as function of their luminosity.
Following an analogy with the FP of the globular clusters, Djorgovski (1995)
suggested a systematic change in the galaxy concentration as a viable
hypothesis. Hjorth \& Madsen (1995) explained part of the tilt of the FP
($\Delta\beta=0.12$) using a galaxy model with a variable central gravitational
potential correlated with the luminosity. Following a similar idea but using a
model directly related with the observations (the S\'ersic (1968) r$^{1/n}$
law), Ciotti et al. (1996) suggested that a suitable variation (by a factor of
$\sim$ 2--4) of the shape parameter $n$  (i.e. a different degree of
concentration) along the FP would be able to explain the tilt. However,
Prugniel \& Simien (1997) (and see also Graham \& Colless (1997) in the
V--band) showed that  the structural nonhomology  contributes only 
$\Delta\beta\sim0.06-0.08$ and  to fully account for the tilt of the FP they
needed to include some contributions from stellar population and rotational
support. These analysis could be  affected by a limited sample  which had a
small range in S\'ersic indices 2$\lesssim$n$\lesssim$5 which is far from
providing the full range of nonhomology observed by other authors with
1$\lesssim$n$\lesssim$10 in the same absolute magnitude range
-22$\lesssim$M$_B$$\lesssim$-15 (see e.g., Graham \& Guzm\'an 2003).

Taking into account the full range of structural nonhomology observed, we
quantify the contribution to the tilt from the structural nonhomology  in the
B band. We show that the structural nonhomology is able to explain
$\approx$3/4 of the tilt in the B--band (i.e.\ $\Delta\beta$=0.2), which, in
combination with the tilt expected from population variations (i.e.\
$\Delta\beta$$\sim$0.1) can explain completely the observed tilt of the FP.

\section{The scalar virial theorem and the FP tilt}

We use the identity
\begin{equation}
L=c_1I_er_e^2
\end{equation}
and the scalar virial theorem for a stationary stellar system
\begin{equation}
GM=c_2\sigma_0^2r_e
\label{cdos}
\end{equation}
with $L$ being the luminosity, $G$ the gravitational constant, $M$ the mass and $c_1$ and $c_2$
structure terms. Any stellar system in
virial equilibrium is then expected to follow the identity:
\begin{equation} r_e=G^{-1}(c_2c_1^{-1})(M/L)^{-1}\sigma_0^2I_e^{-1}.
\label{fp}
\end{equation} With the  definition given in the Introduction for $I_e$, $c_1=2\pi$.
The FP (Dressler et al.  1987; Djorgovski \& Davis 1987) in the
B--band can be fitted by:
\begin{equation}
r_e\propto(\sigma_0^2)^{0.7}I_e^{-0.85}.
\end{equation}
This relation is a consequence of the variation of $c_2^{-1}(M/L)$
with $L$.\footnote{In the more general case, $c_2^{-1}(M/L)$=$L^{\beta}I_e^{\gamma}$,
 $\gamma$ has been found to be compatible with 0 ($\gamma=-0.01\pm0.13$;
Prugniel \& Simien 1996) and in what follows we will assume $\gamma=0$.}
If, following Faber et al. (1987), $c_2$ is constant elliptical galaxies would
be structurally homologous systems and the tilt would have to be explained as a 
systematic variation of mass--to--light (M/L) ratio with luminosity:
$M/L\propto L^{\beta}$, with $\beta$=0.27$\pm$0.08.

As discussed in the Introduction it is unlikely that this tilt can be explained
by stellar population effects alone. One needs to combine two well known
observations: the systematic variation of stellar populations with luminosity
and the change of observed structural properties along the early-type sequence
from dwarf to giant ellipticals.  Ideally both effects in combination would be
able to account for the full tilt of the FP.

\section{The relation between c$_2$ and $n$}

It has been shown (see e.g.  Trujillo, Graham \& Caon 2001 and references
therein) that elliptical galaxies do not form a homologous structural family
and that the luminosity--dependent departures from the r$^{1/4}$ law can be
described by the r$^{1/n}$ S\'ersic model. The nonhomology is also reflected
in the  strong correlations between the shape parameter $n$ and
photometric--independent galaxy properties as, for example, the central
velocity dispersion (Graham, Trujillo \& Caon, 2001).

\begin{figure}

{\includegraphics{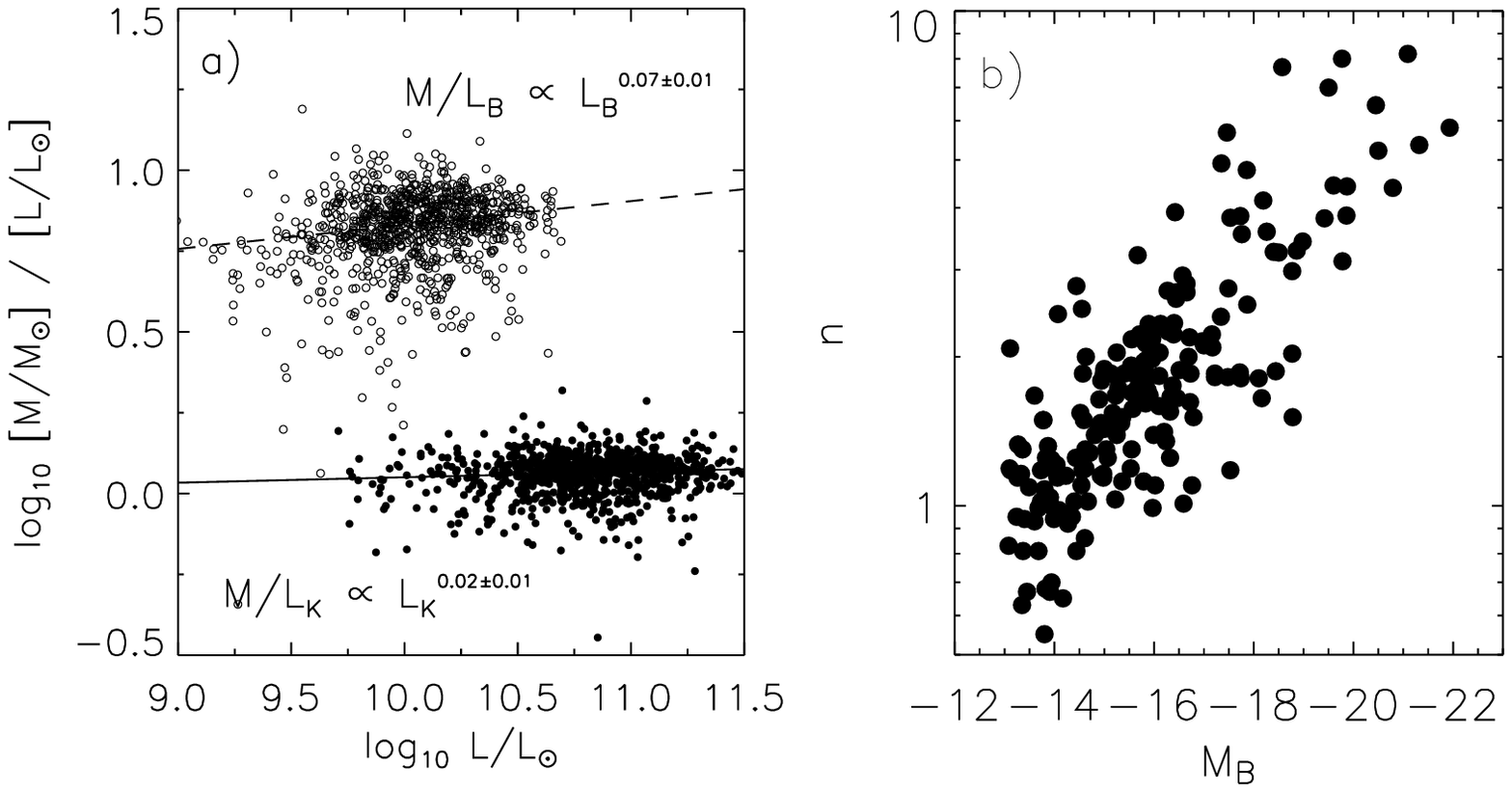}}
\caption[]{({\it a)} Luminosity dependence in stellar M/L ratios in the
$B$-band (open circles) and the $K$-band (filled circles) from 911
morphologically selected early-type  galaxies ($14.5 < r < 16.5;
K < 13.6; r_{e} > 2"$)   from a
combined SDSS/2MASS galaxy catalog. The  stellar M/L ratios are estimated using a
Salpeter (1955) IMF, and using observed galaxy colors to place constraints on
the range of plausible stellar M/L ratios, allowing for generous variations 
in galaxy
stellar population age and metallicity. The $B$-band luminosities and M/L ratios
 were
estimated using stellar population model transformations using the PEGASE code
(see Fioc \& Rocca-Volmerange 1997 for an earlier version of the code that we
use). The lines show robust fits to the trends, where points more than $2.5
\sigma$ from the trends are excluded from the fits. (b)  S\'ersic index $n$
of elliptical  galaxies as a function of their absolute magnitude.}
\label{nohomo}

\end{figure}

In Fig. \ref{nohomo}b, we show the relation between the shape index $n$ and the
absolute B--band (model--independent) magnitude for 200 elliptical galaxies.
The galaxies used in this plot correspond to ellipticals from the Virgo, Fornax
and Coma Clusters (Caon et al. 1990; Caon, Capaccioli \& D'Onofrio 1994;
Binggeli \& Jerjen 1998; Guti\'errez et al. 2004). Those galaxies classified as
S0 were removed from our sample to avoid possible missmeasurements of the index
$n$ due to the disk component of these galaxies. Error estimates for $n$ are
found to have a typical uncertainty of 25\% (Caon, Capaccioli \& D'Onofrio
1993). The high statistical significance of this correlation has been studied
in previous papers (see, e.g. Graham et al. 2001). To check that the robustness
of the above relation is not affected by the large number of faint galaxies,
we have evaluated the Spearman correlation coefficient for those galaxies
brighter than M$_B$=-16.5 (H$_0$=70 km s$^{-1}$ Mpc$^{-1}$) and we find 
r$_s$=0.72.

Using the S\'ersic model,  it is possible to relate the structural
parameter $c_2$ with $n$, leading also to a relation between $n$ and
$L$, or between $c_2$ and $L$. To estimate $c_2$ for a
S\'ersic model we need to evaluate the velocity dispersion
profile $\sigma_r(r)$. In the most simple
case this can be done by assuming a spherical, nonrotating,
isotropic r$^{1/n}$ model and then evaluating $\sigma_r(r)$  using the
Jeans equation (see, e.g., Binney \& Tremaine 1987).
However, the observed central velocity dispersion  $\sigma_0$
that enters in Eq.  \ref{cdos} does not correspond to $\sigma_r(0)$, but rather to the
observed projected velocity dispersion $\sigma_p(R)$, luminosity averaged
over the aperture used for the spectrographic observations
$\sigma_{ap}(R_{ap})$. Consequently, to evaluate properly $c_2$, we
need to switch from $\sigma_r(r)$ to $\sigma_p(R)$ and from this to
$\sigma_{ap}(R_{ap})$ (see a detailed explanation e.g. in Sec. 2.2
of Ciotti et al. 1996).

\begin{figure}
\plotone{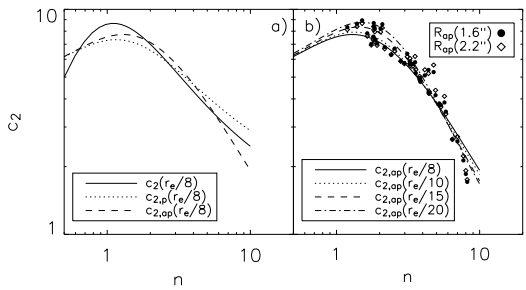}

\caption[]{(a) Estimated virial coefficient $c_2$ for a S\'ersic model based on
an aperture radius $r_e/8$; $c_2$ is obtainted by using the velocity radial
profile $\sigma_r(r)$, the projected radial profile $\sigma_p(R)$ and the
projected aperture radial profile $\sigma_{ap}(R_{ap})$. (b) Estimated  values
of $c_2$ for two different fixed angular apertures (1.$''6$ and 2.$''2$).
Overplotted are the expected $c_2$-values for different apertures.}

\label{cdosplot}
\end{figure}

To illustrate how $c_2$ changes
depending on which estimate of the velocity dispersion is used we
show in Fig. \ref{cdosplot} the relation between $c_2$ and $n$
derived from $\sigma_r(r_e/8)$, $\sigma_p(r_e/8)$ and
$\sigma_{ap}(r_e/8)$. We show $c_2$ for $r=r_e/8$ because this is
one of the most usual aperture radii to measure $\sigma_0$ (J\o
rgensen, Franx \& Kj\ae rgaard 1993). Our value of $c_2$ is in
good agreement with the estimates of this quantity by other
authors (e.g. Prugniel \& Simien (1997), Bertin, Ciotti \& Del
Principe, 2002).

Since the sizes of galaxies differ from one another and because galaxies are
observed at different distances, the estimation of the central velocity
dispersion using a fixed angular aperture samples different fractions of the
total light (or the effective radii). Consequently, when dealing with samples
that extend over a large range of sizes, it is not the best approximation to
estimate $c_2$ from the observations by using a fixed aperture related to the
effective radius (e.g. $r_e$/8). Instead one should use typical angular
apertures (e.g. 1.$''$6 or 2.$''$2).

In order to evaluate the influence of nonhomology in the FP, we
have selected from the sample of elliptical galaxies presented
above those galaxies that have a measured central velocity dispersion.
The velocity dispersions are obtained from Hypercat.
This leaves us with a total of 45 galaxies ranging from -15 to
-22 in the B--band. In Fig. \ref{cdosplot} we show the $c_2$
values for the galaxies of this subsample estimated using
two different fixed realistic apertures 1.$''$6 and 2.$''$2. As we
can see in this figure, the effect of changing the aperture from
1.$''$6 to 2.$''$2 is not dramatic. In what follows we will use
the mean value of these two measurements $c_2(1.''6)$ and
$c_2(2.''2)$ as an estimate of the true value of $c_2$ for our observed
galaxies.

\section{The tilt of the FP}

\begin{figure}
{\includegraphics{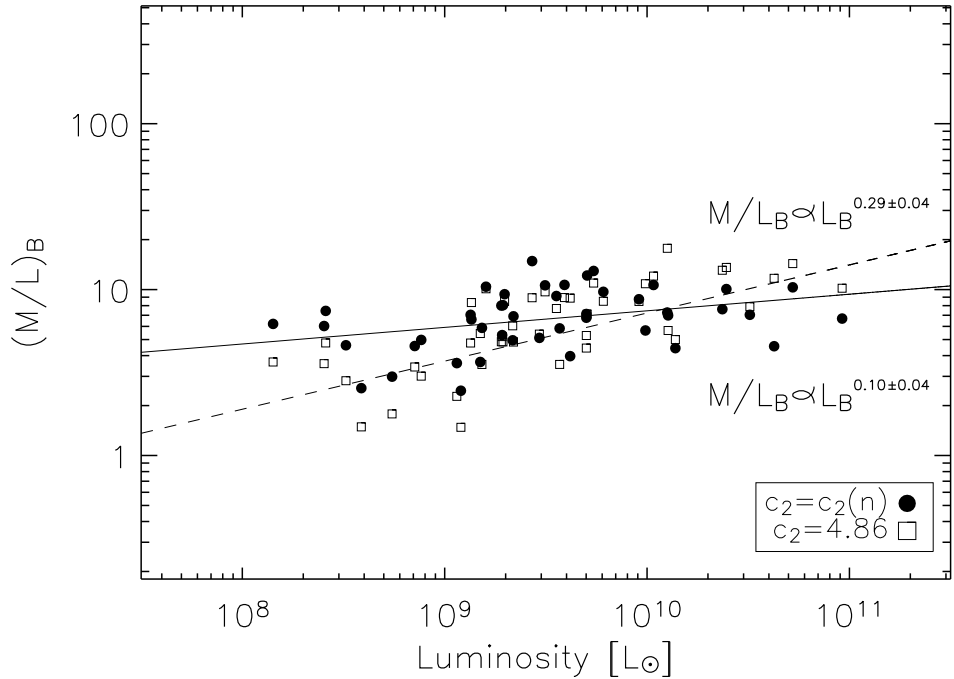}}

\caption[]{Relation between the $M/L$ ratio and $L$ for our selected galaxies.
The filled circles represent the relation between the $M/L$ ratio and $L$ when
$c_2$  accounts for variations in  the observed value of $n$ (i.e. the
nonhomology) of  elliptical galaxies. The open squares represent the relation
between $M/L$ and $L$ when $c_2$ is to assumed to have a constant value (i.e.,
homology) equal to the value used for a King model with $r_t/r_c\sim$100
proposed by BBF. Note that to transform from the units in that
paper to ours, we must make the transformation c$_{2}$=4.3c$_{2,BBF}$.}

\label{tilt}

\end{figure}

Fig. \ref{tilt} shows the relation between $M/L$ and the luminosity $L$ for the
selected galaxies, taking into account the variation in $c_2$ as a result of
variations in shape parameters $n$. For comparison, in the same figure we also
show the result for a constant value for c$_2$, i.e. if structural homology is
assumed to hold. The value we have used for c$_2$=4.86 in the assumption of
homology is taken to match the value presented in BBF for a King model with
$r_t/r_c\sim$100. This value is quite close to the value expected for a de
Vaucouleurs law (n=4; c$_2$=4.6).

Fig. \ref{tilt} shows that the relationship between the $M/L$ and $L$ is much
flatter if the nonhomology is accounted for. To quantify this change we have
performed a minimum $\chi^2$ fit to the data distributions using the relation
$(M/L)=\alpha L^{\beta}$. The values are $\beta=0.10(\pm0.04)$ if nonhomology
is addressed and $\beta=0.29(\pm0.04)$ for the homologous case. The
 exponent $\beta$ for the homologous case in the B-band is in
excellent agreement with the values reported previously. On the other hand, we
can see that the nonhomology part of the tilt accounts for
$\Delta\beta\sim0.2$. We have also checked if the incompleteness is playing a
role at the fainter luminosities. Our sample is 90\% complete down M$_B$=-17.
If we remove from our analysis all the galaxies below this limit we get the
following exponents: $\beta=0.27(\pm0.04)$ for the homologous case and
$\beta=0.06(\pm0.04)$ in the nonhomologous case. Again, we find that the
nonhomology part of the tilt accounts for $\Delta\beta\sim0.2$.

\section{Discussion}

It is encouraging that the tilt $\beta\sim0.1$ of the FP after correction for
to nonhomology  is in very good agreement with the value expected due to
systematic variations of stellar population properties  of elliptical galaxies
($\Delta\beta\sim0.07$). The observed tilt of $\beta=0.29(\pm0.04)$ in the
B--band FP therefore can be understood as a combination of three-quarters
structural nonhomology  and one-quarter stellar population effects. The tilt of
the K--band FP where the change in the stellar population is negligible is
approximately $\beta=0.2$. This is indeed in good agreement with our
expectation if the tilt were due only to the nonhomology and if color
gradients inside galaxies do not change their S\'ersic indices. Indeed, radial
V--K gradients in the surface brightness profiles of ellipticals are on average
only 0.16 mag dex$^{-1}$ (Peletier, Valentijn \& Jameson 1990) which is small
compared to the variations in surface brightness within an effective radius.
The homologous FP of our sample has a scatter in log(M/L) of order 25\% which
is consistent with previous results (BBF; Pahre,
Djorgovski \& Carvalho  1998). The scatter is not significantly reduced in the
non-homologous FP which indicates that it is mainly a result of population
effects and not of structural variations within a given luminosity bin.

Our results rule out that even very massive ellipticals are strongly dominated
by dark matter in their inner regions, which is in agreement with Rix et al.
(1997) and Gerhard et al. (2001). After correcting for nonhomology, the mean
value of the mass-to-light ratio (M/L$_B$) of early-type galaxies in our
luminosity range is 7.1$\pm$2.8 (1$\sigma$ scatter). This value is in good
agreement with the value expected for an old stellar populations of
M/L$_B$=7.8$\pm$2.7 (Gerhard et al 2001).

\acknowledgments

We thank Nicola Caon and Helmut Jerjen for providing us with an (updated)
electronic copy of the Virgo Cluster Catalogs and Hans--Walter Rix, Marijn
Franx, Alister W. Graham and Ralf Bender for interesting discussions. Some data
used in this work were been obtained from HyperLeda database.

\clearpage
\vskip 16.0cm

\clearpage

\clearpage

\end{document}